\documentclass[12pt]{iopart}
\usepackage{iopams,epsf}  
\begin{document}

\title{Thermal lensing in cryogenic sapphire substrates}

\author{Takayuki Tomaru\dag\footnote[7]{Present address: High Energy Accelerator Research Organization (KEK), 1-1 Oho, Tsukuba, Ibaraki, 305-0801, Japan \\
E-mail: tomaru@post.kek.jp},
Toshikazu Suzuki\ddag, Shinji Miyoki\dag, Takashi Uchiyama\ddag, C.T. Taylor\dag, 
Akira Yamamoto\ddag, Takakazu Shintomi\ddag, Masatake Ohashi\dag, Kazuaki Kuroda\dag}

\address{\dag\ Institute for Cosmic Ray Research (ICRR), The University of Tokyo, 5-1-5 Kashiwanoha, Kashiwa, Chiba, 277-8582, Japan}
\address{\ddag\ High Energy Accelerator Research Organization (KEK), 1-1 Oho, Tsukuba, Ibaraki, 305-0801, Japan}

\begin{abstract}
We report the reduction of the thermal lensing in cryogenic sapphire mirrors, which is planed to be used in the Large scale Cryogenic Gravitational wave Telescope (LCGT) project.
We measured three key parameters of sapphire substrate for thermal lensing at cryogenic temperature.
They are optical absorption coefficient, thermal conductivity and temperature coefficient of refractive index at cryogenic temperature.
On basis of these measurements, we estimated the shot noise sensitivity of the interferometer with thermal lensing by using a wave-front tracing simulation. We found that thermal lensing in cryogenic sapphire mirrors is negligible.
\end{abstract}



\maketitle

\section{Introduction}               
Thermal lensing is a serious problem for the next generation interferometric gravitational wave detectors.
Thermal lensing is an effect of wave-front distortion of laser beam caused by refractive index distribution in the mirrors with temperature distribution due to optical absorption.
It causes mode mismatching between beam wave-front and mirror surface, and it finally increases the photon shot noise\cite{fftvirgo2}.
Although mono-crystalline sapphire mirror substrate is one of the most effective materials to reduce thermal noise\cite{saulson,Levin,bragin}, it has large optical absorption at present stage.
Large effort to reduce thermal lensing by reducing optical absorption in sapphire are done by some groups\cite{stan,blair,benabid}.
In this paper, we report the reduction of the thermal lensing in the cryogenic sapphire mirrors, which are planed to be used in the Large scale Cryogenic Gravitational wave Telescope (LCGT) project\cite{lcgt} to reduce thermal noise drastically\cite{cryoQ,fiberQ}.
Firstly, we describe the measurement of key parameters of sapphire substrate at cryogenic temperature, which are needed to estimate wave-front distortion caused by thermal lensing.
After then, we discuss the shot noise sensitivity with thermal lensing by using a wave-front tracing simulation (an FFT simulation)\cite{code}. 

\section{Measurement of key parameters of sapphire substrate for the thermal lensing at cryogenic temperature}
The sagitta of wave-front distortion of laser beam $ds$ caused by the thermal lensing depends on three parameters of mirror substrate;
\begin{equation}
ds \propto \frac{\alpha \beta}{\kappa}
\end{equation}
where $\alpha$ is optical absorption coefficient in the mirror, $\kappa$ is thermal conductivity of the mirror substrate and $\beta$ is temperature coefficient of refractive index ($dn/dT$) of the mirror substrate. 

Firstly, we measured optical absorption coefficient in cryogenic sapphire by laser calorimetry method\cite{lasercal}.
The sample temperature was 5\,K and light wavelength was 1.064\,$\mu$m (Nd:YAG laser).
We measured two sapphire samples, "CSI white high purity" and "Hemlite", both manufactured by Crystal Systems Inc\cite{hem}.
The size of "CSI white" is 10\,mm in diameter and 150\,mm in thickness, and that of "Hemlite" is 100\,mm in diameter and 60\,mm in thickness.
The results show about 90\,ppm/cm optical absorption for both samples\cite{abs}.
There is also no large difference of optical absorption coefficient at different positions in a sample for both samples.
At room temperature, there are several reports for optical absorption in small sapphire samples by photo-thermal method\cite{stan,blair,benabid,schiller}. 
Although their values scatter in a range from 3\,ppm/cm to 550\,ppm/cm, typical values are in between 40\,ppm/cm and 100\,ppm/cm.

Secondly, we measured thermal conductivity of sapphire in a range from 4\,K to 40\,K.
Thermal conductivity of crystals at cryogenic temperature largely depends on sample quality.
The "CSI white" sample, used in optical absorption measurement, was used.
The measurement method is the longitudinal heat flow method\cite{thconddat}.
The result shows that the maximum thermal conductivity of the sample was $4.3\times 10^3$\,W/m/K at 20\,K.
This value is about four times smaller than the value in the data book\cite{thconddat}.
However, this value is two orders of magnitude larger than the value at room temperature.

Lastly, we measured temperature coefficient of refractive index $\beta$ of cryogenic sapphire by measuring the change of refractive angle of light due to the change of sample temperature.
The displacement of transmitted beam $\delta x$ caused by change of refractive index is described as
\begin{equation}
\delta x = \sin(\frac{\pi}{2} - \theta) \frac{d}{\cos\phi} \frac{1}{\cos\phi} \tan\phi \frac{\delta n}{n}, 
\end{equation}
where $d$ is the sample length, 
$\phi$ is the refractive angle,
$n$ is the refractive index of the sample at temperature $T$ and
$\delta n$ is the change of the refractive index.
Injection beam angle $\theta$ of $60$ degrees was used, which gave maximum displacement of transmitted beam for sapphire.
The "Hemlite" sample, 100\,mm in diameter and 60\,mm in thickness, was used.
To investigate thermal deformation of the sample, the displacement of reflected beam from the sample was also monitored.
Although the measured result was limited by thermal deformation of the sample, the upper limit of $|\beta | \leq 9\times 10^{-8}\,\mathrm{K}^{-1}$ was obtained at 1.064\,$\mu$m in an average between 5\,K and 40\,K.
 This value was two orders of magnitude smaller than that at room temperature\cite{dndT300}.
 
 \section{Thermal lensing in interferometer}
 
\begin{table}[htdp]
\caption{Comparison of thermal lensing parameters for fused silica and sapphire.}
\begin{center}
\small
\begin{tabular}{cccc}
\hline
\ \ & Fused silica (300\,K) & Sapphire (300\,K) & Sapphire (20\,K) \\
\hline
$\alpha$\,[ppm/cm] & 2 - 20 & 40 - 100 & 90 \\
$\kappa\,[\mathrm{W/m/K}]$ & 1.4 & 46 & $4.3\times 10^{3}$ \\
$\beta\,\mathrm{[K^{-1}]}$ & $1.4\times 10^{-5}$ & $1.3\times 10^{-5}$ & $\leq | 9\times 10^{-8} |$ \\
$\alpha \beta/\kappa\,\mathrm{[W^{-1}]} \quad \times 10^{-9}$ & 2 - 20 & 1 - 3 & $\leq | 2\times 10^{-4} |$ \\
\hline
\end{tabular}
\end{center}
\label{chfact;ta}
\end{table}%
 
Table \ref{chfact;ta} lists the thermal lensing parameters for fused silica and sapphire.
This table shows that thermal lensing in cryogenic sapphire mirrors is at least four order of magnitude smaller than that both in fused silica and in sapphire at room temperature.
We can regard the thermal lensing in fused silica at room temperature as almost same as that in sapphire at room temperature.

\begin{figure}
\begin{center}
\epsfxsize=80mm
\epsfbox{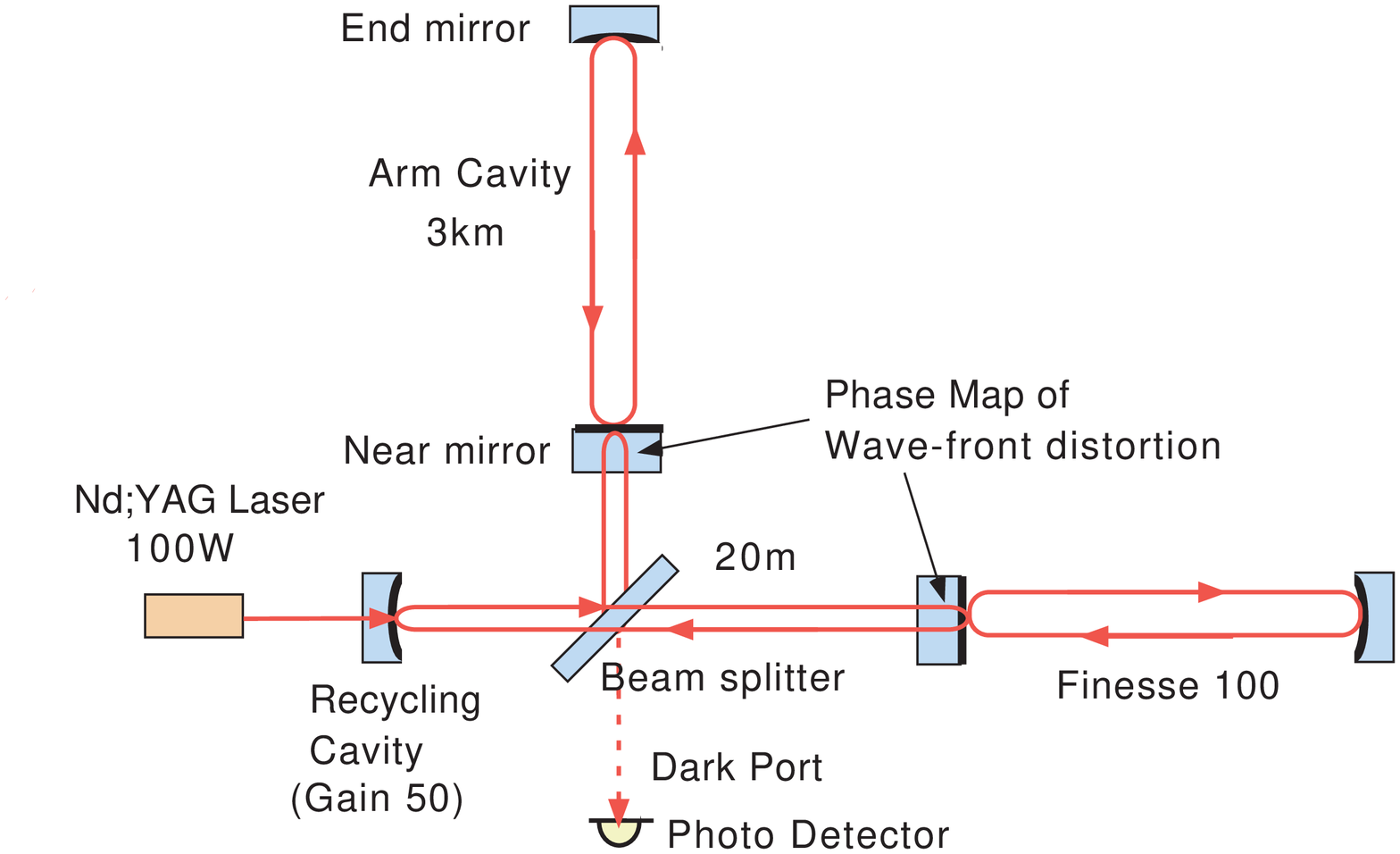}
\caption{A model of GW interferometer (power recycled Fabry-Perot Michelson interferometer).
The phase maps of wave-front distortion caused by thermal lensing were inputted into the both near mirrors.
The LCGT design was used in this calculation.}
\label{RFPM}
\end{center}
\end{figure}

\begin{figure}
\begin{center}
\epsfxsize=65mm
\epsfbox{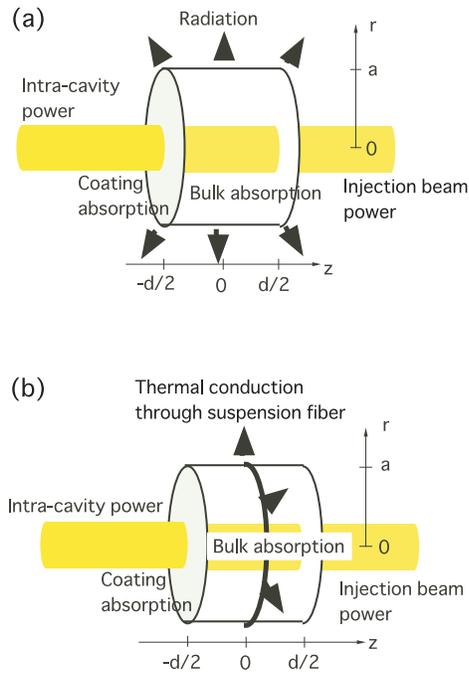}
\caption{A calculation model of wave-front distortion caused by thermal lensing.
(a) Room temperature case (Hello's model). Thermal equilibrium in the mirror is achieved by optical absorption and thermal radiation\cite{calphase}.
(b) Cryogenic temperature case. 20\,K was assumed in the LCGT design.
Thermal equilibrium in the mirror is achieved by optical absorption and thermal conduction of a fiber.
As the first approximation, axi-symmetric heat flow was assumed.}
\label{phase}
\end{center}
\end{figure}

To estimate the shot noise sensitivity of interferometers affected by thermal lensing, an FFT simulator is useful.
The FFT simulator was developed by LIGO\cite{code}.
We assumed optical absorption in sapphire substrate of 90\,ppm/cm and that in coating of 1\,ppm.
The LCGT design, the configuration is a power recycled Fabry-Perot Michelson interferometer (Figure \ref{RFPM}), the finesse is 100, the recycling gain is 50 and the injection laser power is 100\,W, was used in this calculation.
The phase maps of wave-front distortion caused by thermal lensing was calculated by Hello's method\cite{calphase} (Figure \ref{phase} (a)).
In the case of cryogenic mirrors, since thermal radiation from the mirrors is very small and the mirrors are cooled only by thermal conduction of suspension fibers, we calculated the phase maps by changing from the boundary condition of Hello's calculation, which is the thermal radiation, to that of thermal conduction of a fiber (Figure \ref{phase} (b)).
Where we assumed an axi-symmetric model of thermal conduction as the first approximation.
Since the calculation of thermal lensing in the beam splitter has difficulty of axi-asymmetric temperature distribution, we took this effect into the simulation by assuming different thermal lensing between both arms.
 No compensative technique for wave-front distortion\cite{compe1,compe2,compe3} was considered in this calculation.

\begin{figure}
\begin{center}
\epsfxsize=100mm
\epsfbox{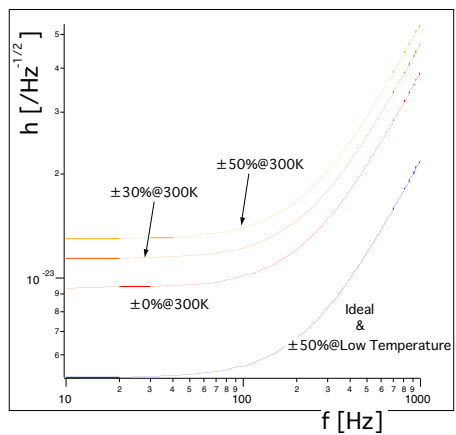}
\caption{Calculation results of the shot noise sensitivity with and without thermal lensing in sapphire mirrors both at room temperature and at cryogenic temperature. 
"Ideal" means the shot noise for the ideal interferometer.
The values of "$\pm 0\%$", "$\pm 30\,\%$" and "$\pm 50\,\%$" mean asymmetric optical absorption between the both near mirrors. }
\label{asym}
\end{center}
\end{figure}

 Figure \ref{asym} shows the results of the shot noise sensitivity calculated by using the FFT simulation.
This result shows that the shot noise sensitivity in the case that  sapphire mirrors are used at room temperature is two times worse than that in the ideal case due to thermal lensing.
This sensitivity decline is resulted from the reduction of recycling gain, from 50 to 16. 
In the case that there are different optical absorption between the both arms, decline of contrast also reduces the sensitivity.
 On the other hand, the thermal lensing in cryogenic sapphire mirrors is negligible, even if the case of $\pm 50$\,\% asymmetric optical absorption between the both near mirrors.
 Cryogenic sapphire mirror effectively prevents thermal lensing.
 
 \section{Summary}
 We estimated the thermal lensing in cryogenic sapphire mirrors on the basis of measurements.
We conclude that cryogenic sapphire mirror is effective for not only reducing thermal noise but also suppressing thermal lensing.
As a remaining problem of cryogenic interferometers, there is a cooling problem due to large optical absorption in sapphire substrate, which will be solved in near future.

\ack
This study was carried out under the MOU between LIGO\cite{ligo} and TAMA\cite{tama}.
We express our appreciation to Dr. B. Bochner, Dr. P. Saha and Dr. Y. Hefetz, the developers of the LIGO's FFT simulator. 
We wish to express appreciation to Dr. H. Yamamoto for preparation of the FFT simulator.
We also express particular thanks to Dr. P. Hello and Dr. J. Y. Vinet, all the developers of the early FFT simulation code.

\end{document}